\documentclass[
    ,final            
  ]
  {aipproc}

\layoutstyle{6x9}

\begin{document}

\title{Fine Tuning in Supersymmetric Models}


\author{\underline{Peter Athron}}{
  address={Department of Physics and Astronomy, 
University of Glasgow,
Glasgow G12 8QQ,
U.K.}
}

\author{D.J.~Miller}{
  address={Department of Physics and Astronomy,
University of Glasgow,
Glasgow G12 8QQ,
U.K.}
}

\begin{abstract}
The solution of a fine tuning problem is one of the principal
motivations of Supersymmetry. However experimental constraints
indicate that many Supersymmetric models are also fine tuned (although
to a much lesser extent). We review the traditional measure of this
fine tuning used in the literature and propose an alternative. We apply this to the MSSM and show the implications.
\end{abstract}

\maketitle



Although Supersymmetry removes fine tuning from the Higgs mass, LEP
constraints may have exposed a MSSM fine tuning problem in the mass of
the $Z$ boson. The minimisation of tuning is now a major motivation in
model building (see e.g.\cite{Kitano:2005wc}).

 Barbieri and Giudice\cite{Barbieri:1987fn} measured tuning in $M_Z^2$,
 with respect to a parameter $p_i$, using a measure
 $\triangle_{BG}(p_i) = \Big{|}\frac{p_i}{O(p_i)}\frac{\partial
 O(p_i)}{\partial p_i}\Big{|}$, giving the percentage change in
 the observable, $O(p_i)$, due to a one percent change in the
 parameter $p_i$. A large $\triangle_{BG}(p_i)$ implies that small
 changes in the parameter result in large changes in the observable,
 so the parameter must be carefully ``tuned'' to the observed
 value. Since there is one $\triangle_{BG}(p_i)$ per parameter, they
 defined the largest to be the tuning for that point, $\triangle_{BG}
 = {\rm max}({\triangle_{BG}(p_i)})$.

However this measure has several problems when applied to complicated
theories.  Firstly a tuning measure should really consider all of the
parameters simultaneously. Also, in attempts to alleviate the tuning
in $M_Z^2$, the problem is often moved into other observables
\cite{Schuster:2005py}. This can be missed when 
only one observable is considered. So a method that can determine both
an individual tuning for each observable and an combined tuning, in
which all of the observables are considered, is desirable.
Additionally, $\triangle_{BG}$ only considers infinitesimal variations
in the parameters. Since MSSM observables are complicated functions of
many parameters, there may be locations where some observables are
stable (unstable) locally, but unstable (stable) over finite
variations. Finally, there is also an implicit assumption that all
values of the parameters in the effective softly broken
Lagrangian ${\cal L}_{SUSY}$ are equally likely, but they have been
written down without knowledge of the high-scale theory, and are
unlikely to match the parameters in the high-scale Lagrangian,
e.g. ${\cal L}_{GUT}$.

\begin{figure}[ht]
    \includegraphics[height=30mm]{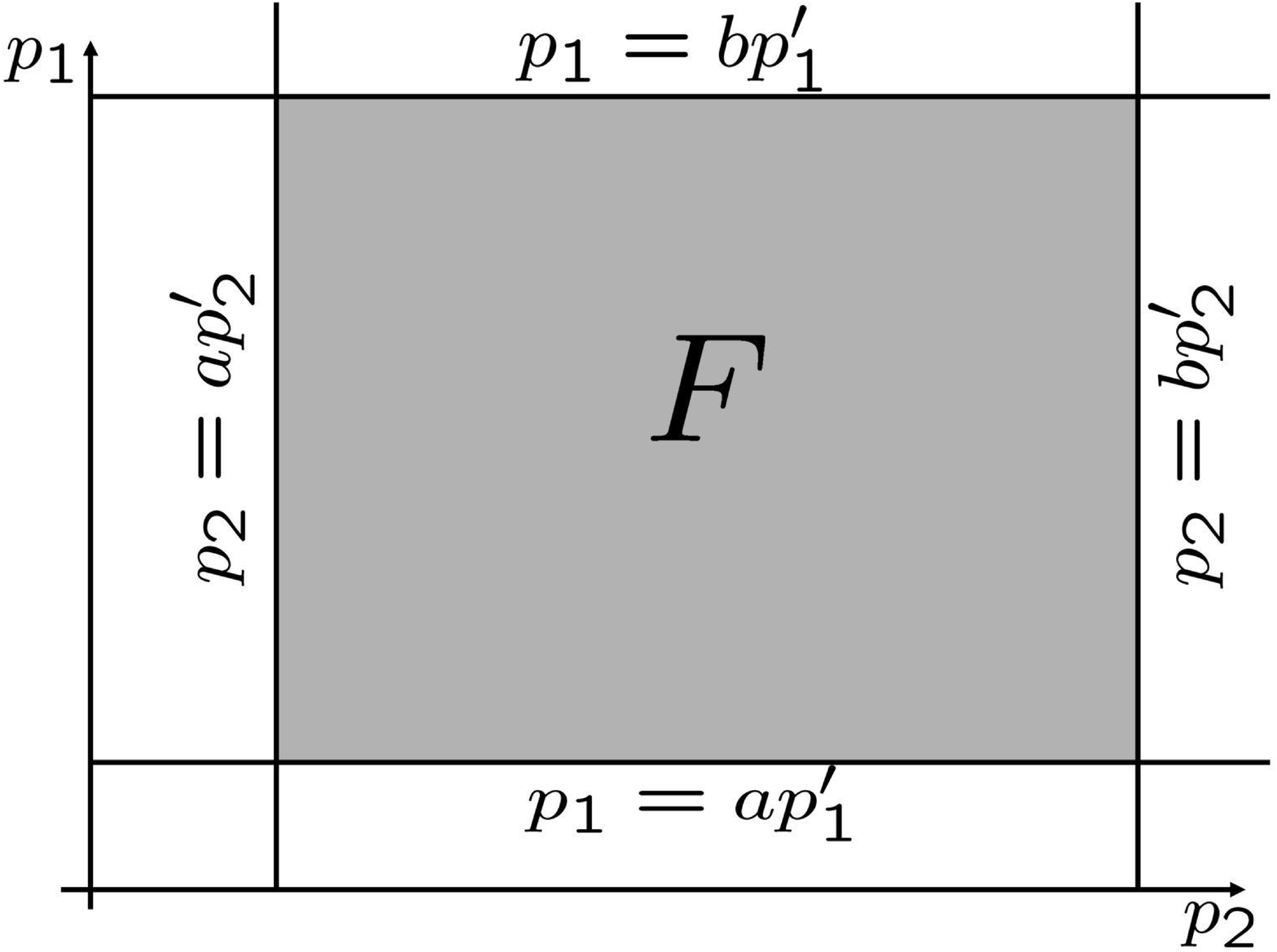}  \hspace*{8mm}
    \includegraphics[height=30mm]{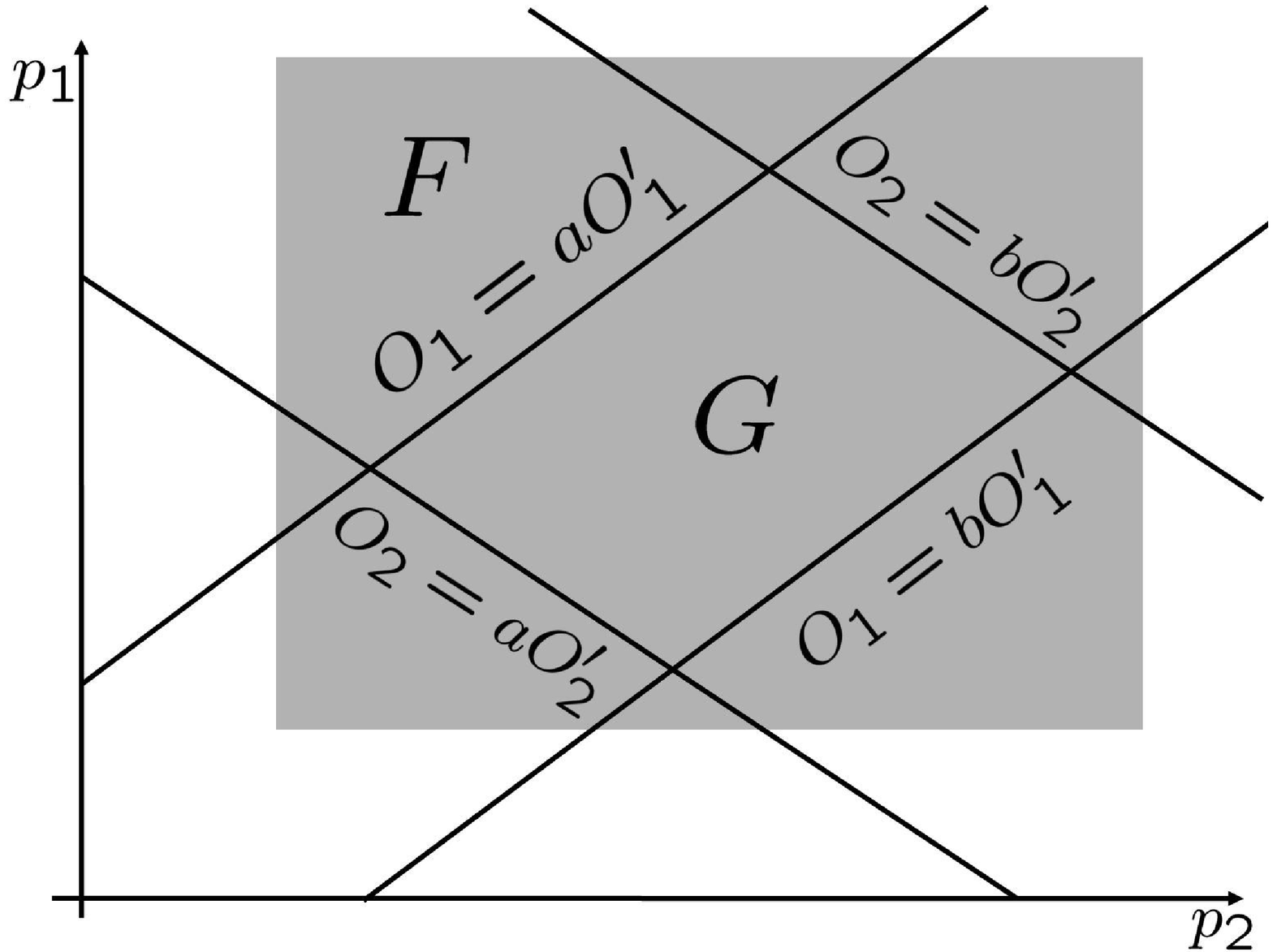} \hspace*{8mm}
    \includegraphics[height=30mm]{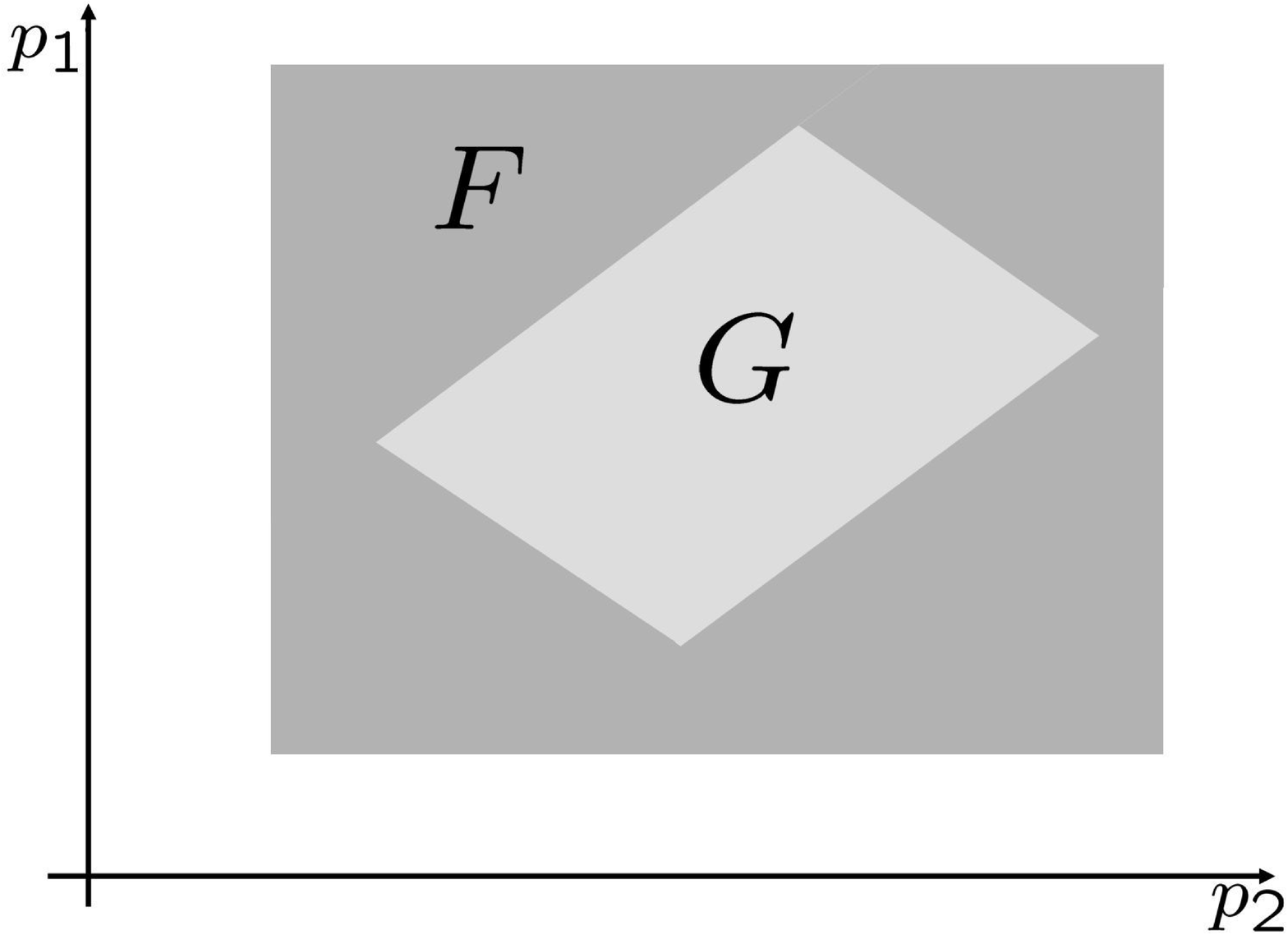}

\caption{Left: In two dimensions the bounds placed on the parameters,
$a\leq\frac{p_i}{p_i'}\leq b$, appear as four lines in parameter space
giving the dark grey area (2d volume), $F$. Middle: Bounds on the two
observables, $a\leq\frac{O_j(\{p_i\})}{O_j(\{p_i'\})}\leq b$ introduce
four more lines which restrict $F$, giving the restricted subset,
volume $G$. Right: Two dimensional volume (area) $F$ (dark grey) and
the restricted subset, area $G$ (light grey). }

\label{Tuning_illustration}
\end{figure}

We propose an alternative measure which accounts for some of these
difficulties.  Tuning occurs when small variations in dimensionless
parameters lead to large variations in dimensionless
observables. For every point $P' = \{p_i'\}$ we define two volumes in
parameter space: $F$ is the volume formed from dimensionless
variations in parameters, $a\leq\frac{p_i}{p_i'}\leq b$, with
arbitrary range $[a,b]$; $G$ is the subset of $F$
restricted to dimensionless variations of the observables,
$a\leq\frac{O_j(\{p_i\})}{O_j(\{p_i'\})}\leq b$. This is illustrated
in Fig.(\ref{Tuning_illustration}). Tuning is then defined by
$\triangle = \frac{F}{G}$.

$\triangle$ quantifies the shrinking of parameter space. This is more
in touch with our intuitive notion of tuning than the stability of the
observable.  With only one or two parameters, $\triangle_{BG}$ also
describes shrinkage in the parameter space and yields the same results
as our new measure. The traditional measure's ability to do this leads
to it's utility as a tuning measure there. Equally it's failure to do
so in many dimensions demonstrates it's limitations.

As a first test of our measure we apply it to the Standard Model Hierarchy Problem.  At one loop we can write, $m^2_h = m^2_0 - C\Lambda^2$, where we treat $\Lambda^2$, the ultraviolet cutoff, and $m_0^2$, the bare mass, as the parameters. $C$ is a positive constant which includes gauge and Yukawa couplings.

\begin{figure}[ht]
\includegraphics[height=50mm]{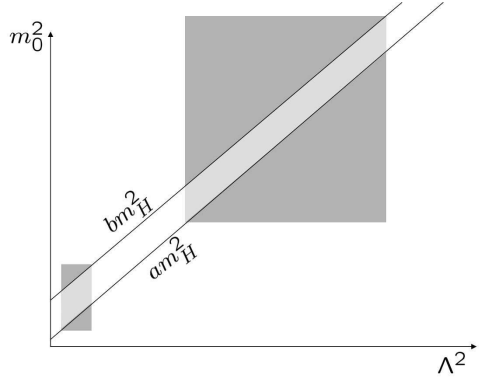}
\caption{The two dimensional volume (area) $F$ (dark grey) and the restricted subset, area $G$ (light grey) for two different points in the two dimensional parameter space.}
\label{2d_SM}
\end{figure}

Varying these parameters about some point $P'(m_0^2, \lambda^2)$ over the dimensionless interval $[a,b]$ defines an area, $F$ in the 2-d parameter space. Applying the same bounds to variations in the Higgs mass gives us two lines in the parameter space. Restricting $F$ to only allow such variations in the observables then defines our area $G$.

This is shown in Fig.(\ref{2d_SM}) for two different points. For one,
the values of the parameters are of the same order as the observable,
$m_h^2$. Here G is not much smaller than $F$. The parameters of the
other point are significantly larger than $m_h^2$. There $F$ is much
larger than $G$, so this point is more tuned than the first. In
general the areas are, $F = (b-a)^2m_0^2\Lambda^2$ and $G
=(b-a)^2\Lambda^2m_h^2$, $\Rightarrow\triangle = \frac{F}{G} = 1 +
\frac{C\Lambda^2}{m_h^2} = \triangle_{BG}$. In this simple case we
find the same result as $\triangle_{BG}$.

In the MSSM there are many parameters, making analytical study
difficult. Here we used a numerical version of our measure.  We take
random dimensionless fluctuations about a point, $P' = \{p_k\}$, to
get $N$ new points $\{P_i\}$. These are passed to a modified version of
Softsusy 2.0.5\cite{Allanach:2001kg} and each point $P_i$ is evolved
from the GUT scale until electroweak symmetry is broken. An iterative
procedure is used to predict $M_Z^2$ and then all the sparticle and
Higgs masses are determined.

As before $F$ is the volume formed by dimensionless variations in the
parameters. $G_{O_i}$ is the sub-volume of $F$ restricted to
$a\leq\frac{O_i(\{p_k\})}{O_i(\{p_k'\})}\leq b$ and $G$ is the
sub-volume of $F$ restricted by
$a\leq\frac{O_j(\{p_k\})}{O_j(\{p_k'\})}\leq b$ for all observables
$O_j$ predicted in Softsusy. For every $O_i$ a count ($N_{O_i}$) is
kept of how often the point lies in the range $G_{O_i}$ as well as an
overall count ($N_O$) of points are in $G$. Tuning is
then measured using, $\triangle_{O_i} \approx \frac{N}{N_{O_i}}$, for
individual observables and $\triangle \approx \frac{N}{N_O}$ for the
overall tuning for that point.

We considered points on the MSUGRA benchmark slope sps1a \cite{Allanach:2002nj},
defined by \mbox{$m_0 = -a_0 = 0.4m_{\frac{1}{2}}$},
$sgn(\mu) = 1$, $\tan{\beta} = 10$. The benchmark point for this slope
has \mbox{$m_0 = 100 {\rm GeV}$}. Table \ref{sps1a_point} shows the
tuning for $M_Z^2$ for two different ranges of variations in the
parameters. Displayed first are tunings obtained by only allowing one
parameter to vary, as for $\triangle_{BG}(p_i)$. Results for R1 and R2
are the same, within statistical errors.

Also shown is our tuning measure for $M_Z^2$, where all parameters
vary simultaneously. Here there is a $2\sigma$ deviation between
results from R1 and R2. This could be a large statistical fluctuation
or actual dependence on the range. However, for individual tunings,
like $\triangle_{M_Z^2}$, a further complication may explain the
deviation.

In particular, using Softsusy to predict the masses for the random
points, sometimes we may have a tachyon, the Higgs potential unbounded
from below, or non-perturbativity. Such points don't belong in volume
$G$ as they will give dramatically different physics. However it is
unclear in which volumes, $G_{O_i}$ (significantly $G_{M_Z^2}$), the
point lies. Such points never register as hits in any of the $G_{O_i}$
and this may artificially inflate the individual tunings, including
$\triangle_{M_Z^2}$. Keeping the range as small as allowed by errors,
minimises the number of problem points.

\begin{table}
\begin{tabular}{lcccc} \hline
            & $\triangle_{M_Z^2}(m_{\frac{1}{2}})$ & $\triangle_{M_Z^2}(\mu)$ & $\triangle_{M_Z^2}(y_t)$ & $\triangle_{M_Z^2}(ALL)$\\ \hline
$R1$   & $38^{+4}_{-4}$ & $41^{+5}_{-5}$ & $25^{+2}_{-2}$ & $56^{+4}_{-4}$ \\ 
$R2$       & $34^{+4}_{-3}$  & $47^{+6}_{-5}$  & $24^{+2}_{-2}$ & $44^{+4}_{-4}$\\ \hline 

\end{tabular}
\caption{Tuning in $M_Z^2$ for sps1a. R1 and R2 are the dimensionless ranges $[\frac{1}{2}, 2]$ and $[0.9, 1.1]$ respectively.}
\label{sps1a_point}
\end{table}

\begin{figure}[h!]
\label{graph}
\resizebox{!}{7cm}{%
    \includegraphics{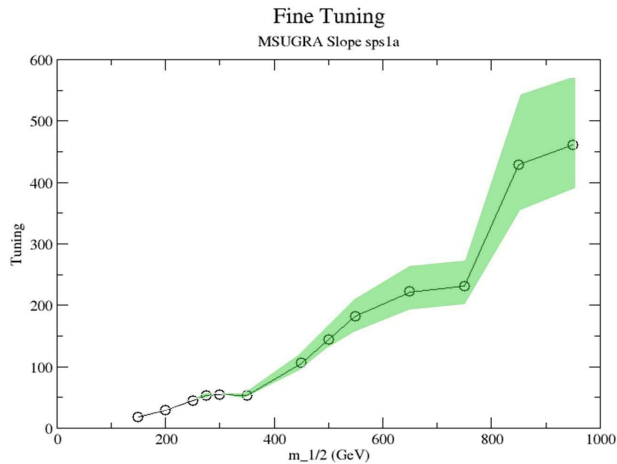}
}
\caption{Tuning in $M_Z^2$ for sps1a slope. Statistical errors (one standard deviation) are shown by the shaded area about the plot.}


\end{figure}

Next, using the smaller range, we applied our tuning measure for
$M_Z^2$ to 12 more points on the sps1a slope. Moving along this slope
in $m_{\frac{1}{2}}$ is an increase in the overall Susy breaking
scale, since the magnitude of every soft breaking term is
increasing. We have plotted the results of this investigation in
Fig.(\ref{graph}).

As expected there is a clear increase in tuning as the Susy breaking
scale is raised. The statistical error, shown by the shaded region,
also increases with tuning, making the numerical approach most
difficult to apply when tuning is large. However precise
determinations of tuning are only relevant for moderate and low
tunings. With tunings greater than $500$, precise values are not a
likely requirement.
 
More work is needed before we can say anything definite about the
overall tuning for the points here. We expect that the dominant tuning
in the MSSM is $M_Z^2$, but nonetheless it will be interesting to see
if there are any other significant contributions.

In summary the SM requires a tuning $\approx 10^{34}$ and this
provides strong motivation for low energy supersymmetry. From searches
at LEP it now appears that the MSSM may also be tuned, though only
$\approx 10^{2}$. If true, rather than being a pathology, this may
provide a hint for a GUT theory. Current measures of tuning cannot
address this question, though, as they neglect the many parameter
nature of fine tuning, ignore additional tunings in other observables,
consider local stability only and assume $ {\cal L}_{SUSY}$ is
parametrised in the same way as ${\cal L}_{GUT}$. Here we have
presented a new measure to address these issues. We have applied this
measure, with uniform probability distributions, to the MSSM
confirming that tuning in the $Z$ Boson mass increases with the Susy
breaking scale.

\end{document}